\documentclass[aps,twocolumn,showpacs]{revtex4}
\usepackage{amsmath}
\usepackage{amssymb}
\usepackage[dvips]{graphicx}
\newcommand{\be}{\begin{equation}}
\newcommand{\ee}{\end{equation}}
\newcommand{\ba}{\begin{eqnarray}}
\newcommand{\ea}{\end{eqnarray}}
\newcommand{\lb}{\label}

\newcommand{\half}{\frac{1}{2}}
\newcommand{\dd}{\partial}

\begin{document}

\title{Riemann-Christoffel flows}

\author{ Patricio S. Letelier}
\email{e-mail: letelier@ime.unicamp.br} 
 
\affiliation{
Departamento de Matem\'atica Aplicada-IMECC,
Universidade Estadual de Campinas,
13081-970 Campinas,  S.P., Brazil}

\begin{abstract}
 A geometric flow based in the Riemann-Christoffel curvature tensor that in two dimensions has some common features with the usual Ricci flow is presented. For $n$ dimensional spaces this new  flow takes into account all  the components of the intrinsic curvature. For four dimensional  Lorentzian manifolds it is found that  the solutions of the Einstein equations associated to a ``detonant" sphere of matter, as well, as a Friedman-Roberson-Walker cosmological  model are examples of Riemann-Christoffel flows. Possible generalizations are mentioned.
\vspace*{0.5cm}

\end{abstract}

\maketitle

The Ricci flow 
\be \frac{\dd g_{ab}}{\dd \lambda}=R_{ab}, \lb{rf} \ee
was introduced by Hamilton \cite{hamilton}  in 1982 and almost immediately applied
to the classification of 3-manifolds. This  work has its  zenith with  the 
celebrated results of Perelman \cite{perelman} that has a notable recognition.  A nice introduction to Ricci flows can be found in \cite{topping}.

Motivated by different mathematical and physical applications the original Ricci
 flow formula suffered many generalizations that can be classified in two groups. 
In the first, new terms are added to the Ricci tensor, e.g., 
$(kR+\Lambda)g_{ab}$ where $k$ and $\Lambda$ are constants and $R$ is the Ricci
 scalar, and/or a symmetric tensor built with 
vector or scalar fields like $\xi_a \xi_b$, $\nabla_a \Phi  \nabla_b \Phi$, and
 $\nabla_a \nabla_b \Phi$ \cite{sever}.
In the second, besides  $ \frac{\dd g_{ab}}{\dd \lambda}$, terms containing 
second derivatives with respect to $\lambda$ are considered  in order to have a hyperbolic or elliptic partial differential equation rather than a parabolic
 one \cite{chin}.

The mathematical applications are mainly concentrated in the study of Ricci 
flows for  metrics of  two or three dimensional spaces. The physical 
applications deal more with four or higher dimensional spaces.

For two and three dimensional manifolds the Ricci tensor 
 characterizes completely the intrinsic curvature of these spaces. Only in these dimensions, once the Ricci tensor is given  one can find all the components of the Riemann-Chistoffel tensor.
Moreover, loosely speaking,  we have that  for large $n$ the number of components of the Ricci tensor grows like $n^2$ whereas the number of components of the the Riemann-Chistoffel tensor increases  like $n^4$.

Thus, when one applies the usual Ricci flow formula, or any of the generalizations above 
mentioned,  for spaces of four or more dimensions we  have that the Ricci 
tensor only carries  partial information about the space intrinsic curvature. 

Motivated by these considerations I shall introduce a new geometric 
 flow based in the 
Riemann-Christoffel curvature tensor instead of the Ricci tensor. Let me
 define the Riemann-Christoffel ``flow" by the expression,
\ba
 \frac{\dd g_{abcd}}{\dd \lambda}&=&R_{abcd}, \lb{rcf} \\
 g_{abcd}&\equiv &g_{ac}g_{bd}-g_{ad}g_{bc}. \lb{gggg}
\ea
The convention used are:  for the Riemann-Christoffel tensor 
$ R^a_{bcd}=\Gamma^a_{bc,d}- \cdots$ and for the Ricci tensor  $R_{bd}=R^a_{bad}$.
Really, equation (\ref{rcf}) does not have the usual form of a usual flow [$\dot{\bf x}=\bf{f}(x)$], but it can be cast as,
\be
\frac{\dd g_{ij}}{\dd \lambda} H^{ij}_{\;\;abcd}=R_{abcd}, \lb{flike}
\ee
with 
\be 
 H^{ij}_{\;\;abcd}=\delta^{(i}_a \delta^{j)}_c g_{bd} +\delta^{(i}_b \delta^{j)}_d g_{ac}-\delta^{(i}_a \delta^{j)}_d g_{bc}-\delta^{(i}_b \delta^{j)}_c g_{ad},
\lb{H}
\ee
where $\delta^{(i}_a \delta^{j)}_b=\half (\delta^{i}_a \delta^{j}_b+\delta^{j}_b \delta^{i}_b).$
A system of equations like (\ref{flike}) may be named a generalized flow.

To compare this  Riemann-Christoffel generalized flow with the usual Ricci flow  let
me first consider the identity:
\ba
H^{ij}_{\;\;ab}&\equiv& H^{ijc}_{\;\;\;\; acb},\\ \lb{h1}
                &=& (n-2)\delta^{(i}_a \delta^{j)}_b +g_{ab} g^{ij}. \lb{h2}
\ea
Now, from (\ref{h2}) and (\ref{flike}) we find,
\be
(n-2)\frac{\dd g_{ab}}{\dd \lambda}+ g_{ab} g^{ij} \frac{\dd g_{ij}}{\dd \lambda}
=R_{ab}. \lb{ric2}
\ee
Since a two dimensional space is always an Einstein space we have that for $n=2$,
equation (\ref{ric2}) reduces to the single scalar equation,
\be
\frac{1}{g}\frac{\dd g}{\dd \lambda}=\half R,
\lb{scalar}
\ee
where $g$ denotes the determinant of the metric, $g_{ij}$. An equivalent  equation can be obtained from (\ref{rf}) modulo the factor one half that can be absorbed redefining the parameter $\lambda$. We note that eq. (\ref{rcf}) in two dimensions gives us only one independent equation that is equivalent to eq. (\ref{scalar})  contrary
 to the Ricci flow that gives us three. To be more specific for the metric,
\be
ds^2=e^{a(x,y,\lambda)}(dx^2+dy^2),   \lb{m21}
\ee
the Riemann-Christoffel flow gives us the single equation,
\be
4 a_{,\lambda}e^a =a_{,xx}+a_{,yy}, \lb{frcm21}
\ee
that,  modulo a redefinition of $\lambda$, is the same well known equation
 obtained from the Ricci flow. In this case, we have for the Ricci flow  that one 
of the equations is identically null and the other two are equal. For the 
other frequently used form of the two dimensional  metric,
\be
ds^2=dx^2 +[A(x,y,\lambda)]^2dy^2, \lb{m22}
\ee
the Riemann-Christoffel flow gives,
\be
2A_{,\lambda}=A_{,xx}. \lb{frc22}
\ee
For the Ricci flow we have only two different equations, the first is exactly the same
 equation (\ref{frc22})  and the second is  $A_{,xx}=0$, that kills the flow.  
Note that, in this case,  the Riemann-Christoffel flow  obeys exactly a linear heat 
equation.

Now let us comeback to the general case, $n\geq 3$, from (\ref{ric2}) and the tensor,
\be 
K^{ab}_{\;\;\; kl}= \frac{1}{n-2}[\delta^{(a}_k \delta^{b)}_l -\frac{1}{2(n-1)} g^{ab}g_{kl}]\lb{k},
\ee
that has the property
\be
H^{ij}_{\;\;\; ab}K^{ab}_{\;\;\; kl}=\delta^{(i}_k \delta^{j)}_l, \lb{hk}
\ee
we find
\be
(n-2)\frac{\dd g_{ab}}{\dd \lambda}=R_{ab}-\frac{1}{2(n-1)}g_{ab}R. \lb{rflike}
\ee
For $n=3$  this last equation is equivalent to the Riemann-Christoffel flow, we have that only for spaces with these dimensions the Ricci tensor has the same number of components than the Riemann-Christoffel curvature  tensor, six.

For spaces with constant curvature, $K$, in any dimensions we have,
\be
R_{abcd}=-Kg_{abcd}. \lb{K}
\ee
Thus the solution of (\ref{rcf}) is 
\be
g_{abcd}(x,\lambda)=e^{-K\lambda}g_{abcd}(x,0)
\ee
Hence, 
\be 
g_{ab}(x,\lambda)=e^{-\half K\lambda}g_{ab}(x,0)
\ee
We have contraction for spaces of positive curvature and
 expansion for negative curvature.

Now I shall study   examples of particular Riemann-Christoffel geometric flows for
four dimensions Lorentzian manifolds. 

Let me consider the particular spherically symmetric metric,
\be ds^2=[h(t,r,\lambda)]^2 dt^2 -dr^2 -r^2(d\vartheta^2+\sin^2\vartheta d\varphi^2).
\lb{sp} \ee
Equation (\ref{rcf}), in this case,  reduces to
\be    2h_{,\lambda}=-h_{,rr}, \;\; 2h_{,\lambda}=-h_{,r}/r, \lb{eqsp}\ee
that has the general solution $h=f(t)(\lambda-r^2)$, where $f$ is an arbitrary function of the indicated argument. If one regards this spacetime as a solution of the Einstein field equations,
\be R_{ab}-\half g_{ab}R=-T_{ab}, \lb{ee}\ee
we find a diagonal energy-momentum tensor  with $T^0_0=0$ and $ T^{1}_{1}=T^{2}_{2}=T^{3}_{3}= p,$
with $p=4/(\lambda- r^2)$.
Therefore we have a perfect ``fluid" with  pressure  for $\lambda>r^2$  or tension  when $\lambda<r^2$, and no energy density. This fluid may be called a ``detonant fluid", since in a detonation the pressure is by far more important than the  density and this last may be disregarded. We note a dilution of the pressure as $\lambda$  increases. We have a sort of phase transition in the shell of radius $r_c=\sqrt{\lambda}$. In this case the parameter $\lambda$ may  be associated to a thermodynamic  variable.

Now, I shall consider Riemann-Christoffel geometrical  flows for  Friedman-Roberson-Walker  cosmological model 
 \be
ds^2=a^{-1}(t,\lambda)dt^2-a(t,\lambda)(dx^2+dy^2+dz^2) \lb{cm}.
\ee
Equations (\ref{rcf}) give us,
\be
8a_{,\lambda}=  a_{,t}^2, \;\;  a_{,tt}=0, \lb{eqcm}
\ee
whose solution is $a = kt +k^2\lambda /8$, where $k$ is an arbitrary constant. From the Einstein equations we find that the source of this space time is  a perfect fluid with density $\rho= 6k/(k\lambda+8t)$  and pressure $p=\rho/3$, i.e., we have an expanding  universe filled by a photon gas.  We have a dilution of the pressure and  density when $\lambda$ increases.

In summary, the Riemann-Christoffel geometric flow (\ref{rcf}), in two dimensions has some common features with the usual Ricci flow (\ref{rf}). For $n$ dimensional spaces this flows take into account all  the component of the intrinsic curvature. For four dimensional  Lorentzian manifolds I find that the solutions of the Einstein equations associated to  a very simple ``detonant sphere" of matter, as well as, a flat Friedman-Robertson-Walker cosmological with $p=\rho/3$ equation of state  are examples of Riemann-Christoffel geometric flows. Incidentally, these two examples do not satisfy the Ricci flow equation (\ref{rf}). For manifolds  of dimension $n\geq 4$ a similar flow can be constructed with the Weyl tensor, $W_{abcd}$, that has no meaning for $n < 4$. For $n=4$ the Levi-Civita totally anti-symmetric tensor, $\varepsilon_{abcd}$, can be used to add new terms to  the Riemann-Christoffel flow. These new terms may  involve new fields or other geometrical quantities e.g., 
$\varepsilon_{abcd}(kR+\Lambda)$. Also extension of the Riemann-Christoffel geometric flow involving second derivatives of the 
parameter $\lambda$ may be considered.

\acknowledgements

 I want to  thank R. Mosna,  C. Negreiros,  S.R. Oliveira, 
 and A. Saa for discussions about Ricci flows and also FAPESP and CNPq for 
 partial financial
support. 


\end{document}